\documentclass[preprint,aps]{revtex4}

\usepackage[pdftex]{graphicx}
\usepackage{amssymb,amsfonts,amsmath}

\begin{document}

\title{Atomic-detailed milestones along the folding trajectory of protein G}

\author{C. Camilloni$^1$, G. Tiana$^{1,2}$ and R. A. Broglia$^{1,2,3}$}
\affiliation{$^1$Department of Physics, University of Milano, via Celoria 16, 20133 Milan, Italy.\\
$^2$INFN, Milan Section, Milan, Italy\\$^3$The Niels Bohr Institute, University of Copenhagen, Blegdamsvej 17, DK-2100 Copenhagen, Denmark}


\begin{abstract}
The high computational cost of carrying out molecular dynamics simulations of even small--size proteins is a major obstacle in the study, at atomic detail and in explicit solvent, of the physical mechanism which is at the basis of the folding of proteins. Making use of a biasing algorithm, based on the principle of the ratchet--and--pawl, we have been able to calculate eight folding trajectories (to an RMSD between 1.2\AA~and 2.5\AA) of the B1 domain of protein G in explicit solvent without the need of high--performance computing. The simulations show that in the denatured state there is a complex network of cause-effect relationships among contacts, which results in a rather hierarchical folding mechanism. The network displays few local and nonlocal native contacts which are cause of most of the others, in agreement with the NOE signals obtained in mildly-denatured conditions. Also nonnative contacts play an active role in the folding kinetics. The set of conformations corresponding to the transition state display $\varphi$--values with a correlation coefficient of 0.69 with the experimental ones. They are structurally quite homogeneous and topologically nativeÐlike, although some of the side chains and most of the hydrogen bonds are not in place. \end{abstract}
\maketitle



Molecular--dynamics simulations in explicit solvent can be a very useful complement to experimental studies of protein folding, in keeping with the fact that they provide insight into the time evolution of the process with atomic detail, under fully controlled conditions~\cite{Han:02}. 
On the other hand, they are computationally very demanding, even in the case of small proteins. Among the most massive folding simulations ever realized is a 10$\mu$s molecular dynamics (MD) folding trajectory of the 38--residue WW domain, lasting for about 3 months on 329 cores and reaching conformations which are $\sim 50\%$ similar to the native conformations in terms of number of contacts \cite{Fre:08}.  To be statistically sound, Pande and coworkers carried out 410 simulations of the folding of the 35--residue Villin Headpiece, the average duration being 863 ns. The calculation lasted for 54 machine years on a distributed computer, and eighteen of these trajectories reached the native conformation \cite{Ens:07}.  

The intrinsic and unavoidable computational problem in carrying out folding simulations with realistic protein models is the wide range of time scales involved: the time step of the simulation must be tuned to femtoseconds, corresponding to the time scale of atomic vibrations, while the overall folding process spans over interval of time ranging from milliseconds to seconds. In an attempt to overcome this difficulty, a number of investigations focused on the study of unfolding simulations at high temperature \cite{Bro:98a, Dag:01}.

A decade ago Marchi and Ballone developed an adiabatic bias molecular dynamics (ABMD) method  \cite{Mar:99}, to generate MD trajectories between pairs of points in the conformational space of complex systems. It was applied for the first time to protein unfolding by Paci and Karplus~\cite{Pac:99}. The method is based on the introduction of a biasing potential which is zero when the system is moving towards the desired arrival point and which damps the fluctuations when the system attempts at moving in the opposite direction. As in the case of the ratchet and pawl system, propelled by thermal motion of the solvent molecules, the biasing potential does not exert work on the system. Consequently, the resulting trajectories are physically correct. On the other hand, the algorithm cannot provide the statistical weight of the visited states nor the time scales associated with the trajectory.

In the present work we report on results of the application of the ABMD algorithm to the study, with the help of the Amber force field \cite{Dua:03} in explicit solvent and without recurring to high--performance computing, of the folding of the 56--residues B1 domain of protein G starting from 16 thermally--unfolded conformations. From the eight trajectories which reached a RMSD lower than $2.5$\AA~we have extracted the conditional probabilities of contact formation between the amino acids of the protein. From them it is possible to learn whether there are obligatory steps along the folding pathway of the protein. 


\section{Results}

\begin{figure}
\centerline{\includegraphics[height=7cm]{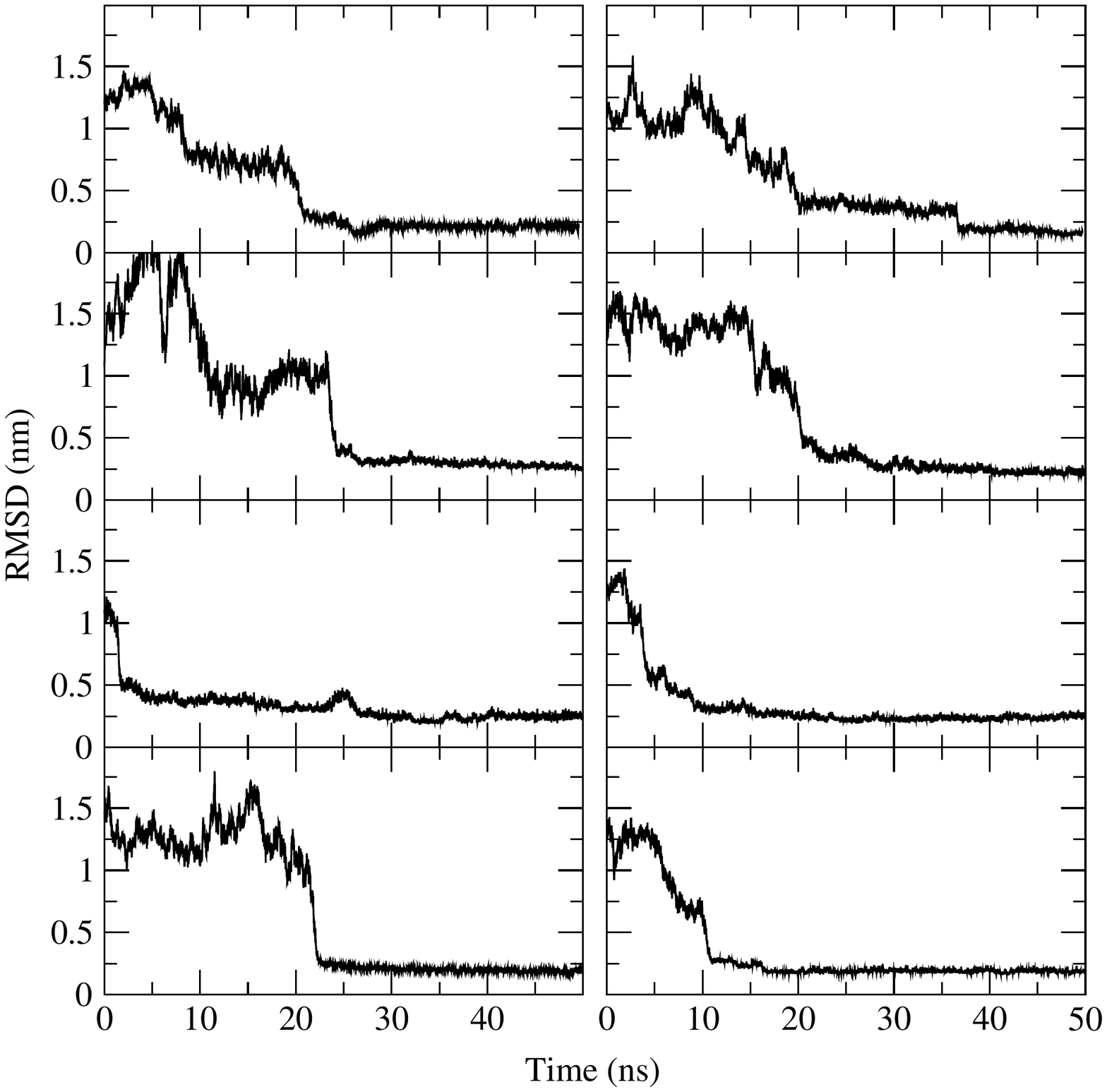}}
\caption{The RMSD calculated on the C$\alpha$ atoms along the 8 folded trajectories}\label{rmsd}
\end{figure}

Sixteen ABMD simulations were carried out starting from uncorrelated high--temperature protein conformations in water, drived by the distance $d_{CM}$ of the contact map of a given protein conformation from the native contact map (cf. Eqs. \ref{eq:dcm1},\ref{eq:dcm2}). Eight of these simulations fold within an RMSD of $2.5$\AA~from the crystallographic conformation within the simulated time of 50 ns  (see Fig.~\ref{rmsd}). Two of the folding trajectories reach conformations within an RMSD of $1.2$\AA, and other two within $1.4$\AA. The other four folding trajectories display imperfections due to non--native alignment of some side--chains, a result which is fostered by the biasing algorithm and which is anyway compatible with the predicted glassy dynamics of side--chains in the native state \cite{Kus:02}. These imperfections cause the RMSD to reach values up to $2.5$\AA, even if the overall topology was correct.

Concerning the eight non--folding trajectories, three of them display the two hairpins docked on the wrong side with respect to the helix; two trajectories display the hairpins undocked but with the orientation of the side chains which is symmetrical with respect to the native conformation. These misfolded conformations are reached because the definition of the reaction coordinate $d_{CM}$ employed to drive the simulation involves mainly the $C_\alpha$ atoms (which are 56) and only 12 atoms belonging to the hydrophobic side chains. Consequently, it is not always effective in discouraging the formation of conformations with a wrong symmetry of the side chains. The structure of the protein in the remaining three trajectories does not display misfolded features and seems only to need longer simulation times to reach the native state. 

\subsection{The kinetics of contact formation is rather hierarchic}
The only information which ABMD simulations can provide concerns the sequence of events of the folding process (which follows what). The order of formation of the 110 native contacts of the protein are calculated for each trajectory, defining the quantity $t(i,k)$ as the (nominal) time at which the $i$th contact is stably formed in the $k$th simulation. From it one can define the probability 
\begin{equation}
M_{ij} = \frac{1}{8}\sum_{k=1}^8 \theta\left(t(i,k)-t(j,k)\right),
\label{eq1}
\end{equation}
that the $i$th contact is formed before $j$, where $\theta$ is the Heaviside's step function. This matrix satisfies $M_{ij}+M_{ji}=1$ and each element $M_{ij}$  assumes the value $1$ if the formation of the $i$th contact precedes the formation of the $j$th, $0$ if it follows it, and $1/2$ if the two are uncorrelated. A related quantity to $M_{ij}$ is the probability $A_j=\sum_{i\neq j} M_{ij}/109$ that the $j$th contact is formed after any other contact. 

The plot of the $A_j$, ordered from the smallest to the largest values (cf. Fig. S1 in the Supplementary Materials), can be used to study to which extent the formation of contacts during folding is hierarchic. If, during protein folding, native contacts are formed along a deterministic hierarchy of events, as in a chain of chemical reactions, the ordered $A_j$ values will lay on the diagonal of the plot. On the contrary, if folding is fully cooperative, in the sense that all native contacts are formed simultaneously at the transition state, the ordered  $A_j$ values will lay on a horizontal line. One can thus define a parameter $hi$ to measure the degree of ``hierarchicity" of the folding process, proportional to the angular coefficient of the ordered $A_j$ values, ascribing to it the value $1$ in the case of a deterministic hierarchy. One should notice that a hierarchical folding kinetics is not incompatible with a sharp thermodynamic transition from denatured to native state at equilibrium, something which can be produced, for example, by a large free-energy barrier at one step of the hierarchy. 

The value of $hi$ associated with the simulation is 0.65, indicating a fairly hierarchical process. One should notice that a large value of $hi$ is not in disagreement with the sharpness of the thermodynamic transition between the denatured and the native state, as this transition usually involves a small subset of native contacts in keeping with the fact that a number of these contacts are already formed in the denatured state \cite{Bro:08}. As a control, a random matrix satisfying the $M_{ij}+M_{ji}=1$ requirement displays $hi=0.08$. A further control case is that of homopolymeric chains whose rate of contact formation depends only on the distance along the chain of the two residues involved (see Supplementary Materials). This model provides a $hi=0.27$, reflecting the hierarchy arising from the straightforward fact that residues which are close along the chain build out contacts faster than those which are far apart along it. 

The curve associated with the folding simulations display six contacts with particularly--low value of $A_j$. These contacts are between pairs of residue 22--25, 29--32, 34--38 and 35--38 (within the helix), 46--49 and 43--54 (within the second hairpin) and 7-54 (between the N-- and the C-terminal strands of the protein). Because the presence of such an early non--local contact is entropically quite unfavorable, its presence underscores its essential role in the whole folding process, strongly restricting the possible conformations of the chains. It is not stabilized by hydrogen bonds or salt bridges, but takes place between two hydrophobic residues (L7 and V54), L7 being close to other two hydrophobic residues (L5 and I6) and belonging to the eventual hydrophobic core of the protein G.

The contacts displaying the largest probability of being formed after all the others have done so, belong mostly to the interface between the two hairpins and between each of them and the helix.  Interestingly, three of these late--forming contacts are between residues stabilizing the first hairpin, namely 9--13, 7--16 and 4--15. 

\subsection{The folding hierarchy involves three levels of contact formation} 
\begin{figure}
\centerline{\includegraphics[height=7cm]{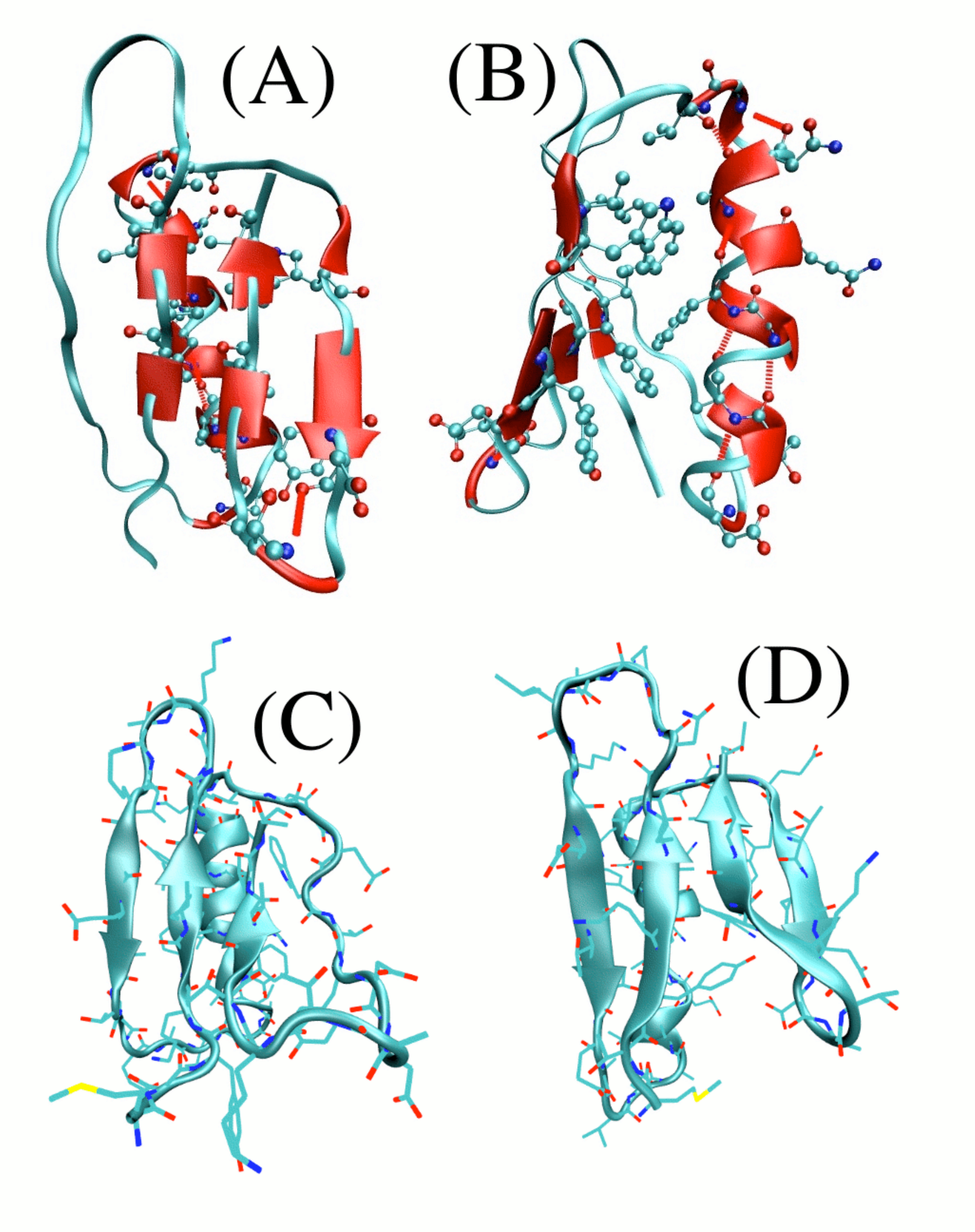}}
\caption{(A) and (B) The  native conformation displayed from two different points of view, with the residues building contacts which are not the consequence of any other contact displayed as cartoons, and their sidechain drawn with balls and sticks. (C) and (D) The transition states obtained from the first two trajectories, respectively.}
\label{strutture}
\end{figure}

To inspect the hierarchy of contact formation, we report in Fig. \ref{freccie} the set of native contacts with respect to the probability $A_j$ that the contact follows the other contacts (and should not be confused with a temporal axis, as the present simulation cannot give information in this respect). Thirteen contacts, boxed with a red square in the figure, are not the consequence of the formation of any other contact, but are themselves the cause of the formation of other contacts (i.e., are contacts $i$ such that $M_{ij}=1$ for all $j$'s, and $M_{ji}=0$ for some $i$). The residues building such contacts are highlighted in red in Figs. \ref{strutture}(A) and (B). They are mostly local contacts within the second hairpin and within the helix. Exception is made for contact 39--54 between the helix and the second hairpin and the non--local contacts 5--30 and 7--54 between the first hairpin and the helix and between the two hairpins, respectively.

The contacts boxed in blue in Fig. \ref{freccie} are those which always follow other contacts and never cause them. They involve all secondary and tertiary  structures of the protein, indicating that no part of the protein has completely reached its native conformation before the overall folding. Essentially each of the red contacts is linked to some blue contact by a cause--effect relationship (i.e., $M_{ij}=1$), indicating that there are two well-defined layers in the hierarchy of contact formation.

The five contacts squared in black constitute an intermediate layer which always follow the formation of contacts 22--25 and 29--32 within the helix, and are themselves cause of the formation of many contacts distributed throughout the protein (the gray arrows join these contacts $j$ with the contacts $i$ such that $M_{ij}=1$ or $M_{ji}=1$). 

The contacts not marked by squares are those displaying fractional probabilities to be the cause or the consequence of the formation of some other contact. These contacts are mostly concentrated within the first hairpin, and to a lesser extent between the two hairpins.

Summing up, the simulation indicates that the folding mechanism of protein G involves first the spontaneous formation of native contacts in the second hairpin, in the helix and few non--local contacts. The stabilization of the first hairpin and the formation of most tertiary contacts come only as a consequence of these events.

\subsection{A small number of non--native contacts are formed with probablity one}
Operatively, a non--native contact is assumed established when two amino acids lying more than three residues apart along the chain and further away than 5\AA~in the native conformation found themselves come, in the folding process, closer than 4\AA.
There are 9 non--native contacts which are formed for some time in all the folding trajectories; these contacts display well--defined behaviours. The contacts E19--A26 and V21--A26 stabilize a non--native turn (hydrophobic staple motif \cite{Ser:97,Bro:98b,Cam:08}) in the region between the first $\beta$--hairpin and the $\alpha$--helix; this turn is disrupted when the first N--terminal turn of the helix is formed. Residues K31 and D40 form a non--native salt--bridge which stabilizes the C-terminal segment of the helix. When the helix forms its tertiary interactions with the hairpin, the salt bridge is broken and the side chain of K31 gets reoriented to interact with the hairpin.

Contacts Y33--D40 and N35--G41 are always formed in the initial collapse of the chain. Contacts V39--T55 and D40--T55 have a bizarre behaviour. They are formed after the C-terminal part of the helix, when no other native contact between the helix and the second hairpin are formed. Their disruption is followed by the formation of the native contacts in the immediate vicinity (e.g. 39--56, 40--56), which substitute such non--native contacts in the docking between the helix and the second hairpin. Consequently, they seem to act as baits to entice together the two secondary structures of the protein to come togheter.

\subsection{The transition state ensemble is structurally homogeneous and displays a very native--like topology}

The ABMD simulations are not able to directly identify where the transition state (TS) lies. On the other hand, since protein folding is associated with the crossing of a free-energy barrier, one expects the TS to be associated with a marked jump in the reaction coordinate of the system. While the correct reaction coordinate is unknown, $d_{CM}$ has proven effective in ratcheting the protein to its native conformation, thus showing to be correlated with it. Consequently, the assumption is made that the transition state is located close to the last jump in $d_{CM}$ before the native state (cf. Fig. S2 in the Supplementary Materials).

\begin{figure}
\centerline{\includegraphics[height=7cm]{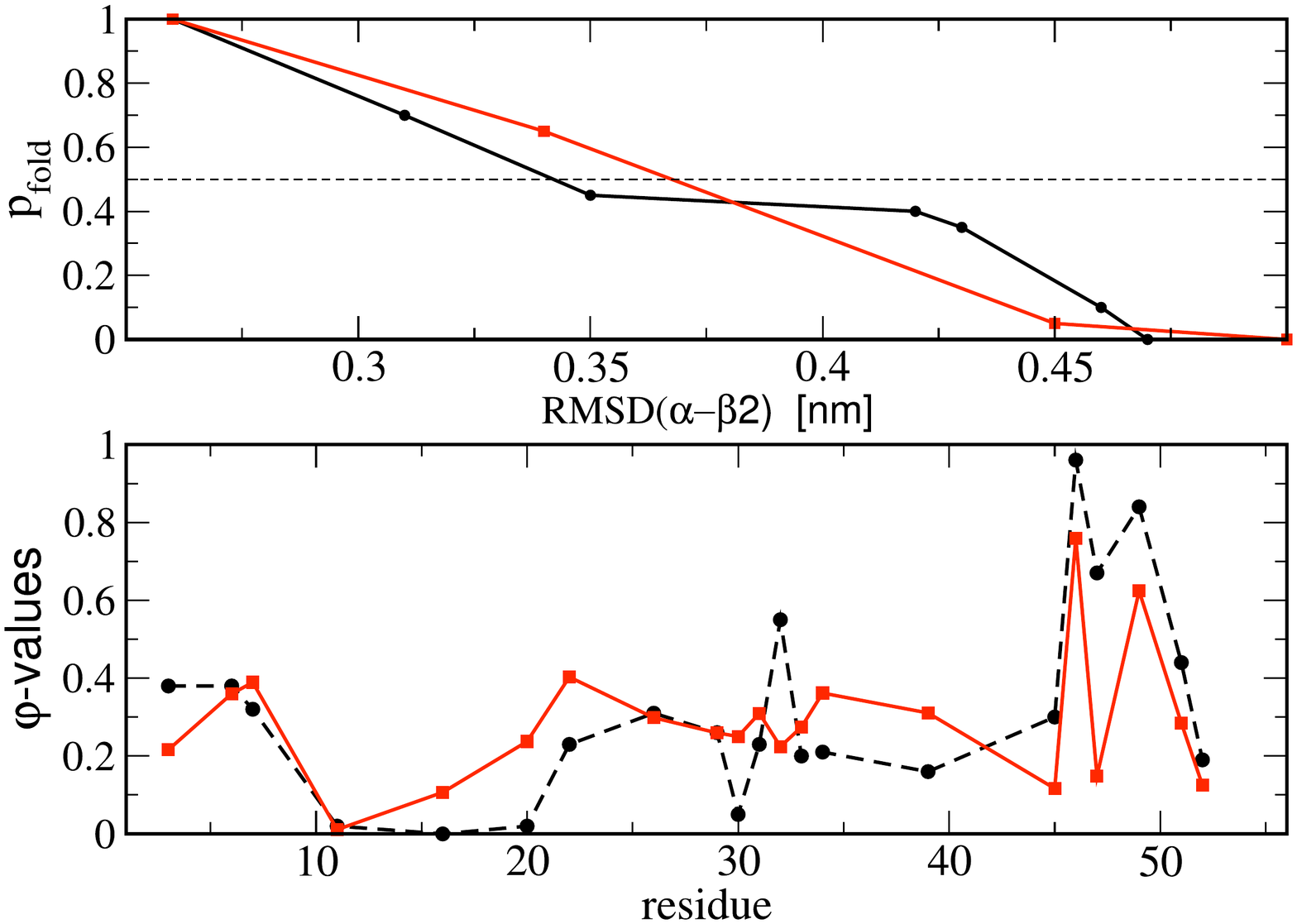}}
\caption{(upper panel) The folding probability of MD simulations starting from conformations with different values of the RMSD between the helix and the second hairpin obtained by two different ABMD trajectories.  The transition state corresponds to the conformations with folding probability equal to 0.5 (dashed line). The RMSD between the helix and the second hairpin has been chosen because results to be the region of the protein that is the least structured in the transition state (33\% of native contacts formed, see text). (left panel, below) The $\varphi$--values obtained by simulations (dashed curve) and by experiments (solid curve).}
\label{phivalues}
\end{figure}

To test the validity of this assumption we have carried out a commitment analysis \cite{Gei:99} on the first two folding trajectories. Starting from (twelve, 7 from the first trajectory and 5 from the second, respectively) different conformations chosen in the neighbourhood of the putative transition state, 20 MD simulations have been carried out for 10 ns each (see Materials and Methods). These calculations corresponds overall to 2.4$\mu s$ of MD. The fraction $p_{fold}$ of folding simulations as a function of the RMSD of the $\alpha$--helix and of the second $\beta$--hairpin is displayed in Fig. \ref{phivalues} and shows a monotonic behaviour, as expected for the transition state. The conformations displaying $p_{fold}=0.5$ correspond, by definition, to the transition state. The most representative among the selected as initial TS conformations for each of the first two trajectories (corresponding to times 20.820 ns and 37.360 ns, respectively) are displayed in Fig. \ref{strutture}(C) and (D).

The TS conformations found in the two independent runs are quite similar to each other, displaying 54\% of common formed native contacts. This result suggests that the transition--state ensemble is structurally rather homogeneous. Moreover, their topologies are very native--like, the RMSD calculated on the $C_\alpha$ atoms being 3.6\AA~and 3.0\AA, respectively. Looking at the secondary and tertiary structures in more detail reveals less ordered features; in fact, the fraction of native contacts in the two TS conformations, defined as in the preceding Sections, are 36\% and 62\%, respectively. The contacts within secondary structures are well formed (81\% of the first hairpin, 94\% of the helix and 69\% of the second hairpin), while those stabilizing tertiary structures are less so (51\% of contacts $\alpha-\beta1$, 33\% of contacts $\alpha-\beta2$ and 57\% of contacts $\beta1-\beta2$). The missing native contacts are mainly the hydrogen bonds associated with the hairpins (both intra-- and inter--hairpin) and the contacts between the side chains of the helix and the two hairpins. In other words, the overall topology is native--like, while the detailed geometry of secondary structures and the packing of the buried side chains is not yet optimized.

The contacts which are not formed in both the transition state conformations are displayed in Fig. \ref{freccie} within dotted ellipses. Most of them belong to the set of contacts  which always follow other contacts and never cause them (marked in blue in the figure), except for contact 28--32.

\subsection{Calculation of $\varphi$--values from the folding trajectories}

The knowledge of the structure of the protein in the transition state allows to calculate the associated $\varphi$--values and thus be able to make a comparison with the experimental findings \cite{paci02}. The $\varphi$--values calculated for the two transition--state conformations displayed in Figs. \ref{strutture}C and D have correlations with the experimental data of ref. \cite{McC:00} of only 0.25.

Indeed, experimental $\varphi$--values consist of an average over a very large number of molecules. Although the transition state obtained from ABMD simulations is topologically quite homogeneous, $\varphi$--values are quite sensitive to the detailed atomic environment of the mutated side chain, resulting in a nontrivial distribution. Using as transition state ensemble the two transition--state conformations  and the six putative conformations obtained in the previous Section, an average set of  $\varphi$--values was calculated and displayed in the lower panel of Fig. \ref{phivalues}, in comparison with the experimental values. The correlation coefficient is now 0.69.
Of notice that these results shed light on the power of the $\varphi$--value analysis of folding, not only as a valuable characterization of the transition state, but also as the specific probe, at the level of single atoms, of the conformations associated with the top of the folding free energy barrier.

\section{Discussion}

With the help of the ABMD algorithm it is possible to simulate, in few days on a PC, the folding of a small--size protein (as e.g. protein G) with atomic detail and in explicit solvent making use of realistic force-fields. One can generate multiple folding trajectories, and consequently obtain a statistically representative information concerning the kinetics of the folding process. The price to be paid for reaching this goal, is that only the simulated succession of events is physically meaningful, while the statistical weights of the visited states and the time scales obtained from the simulations are not.  

\begin{figure*}
\centerline{\includegraphics[height=9cm]{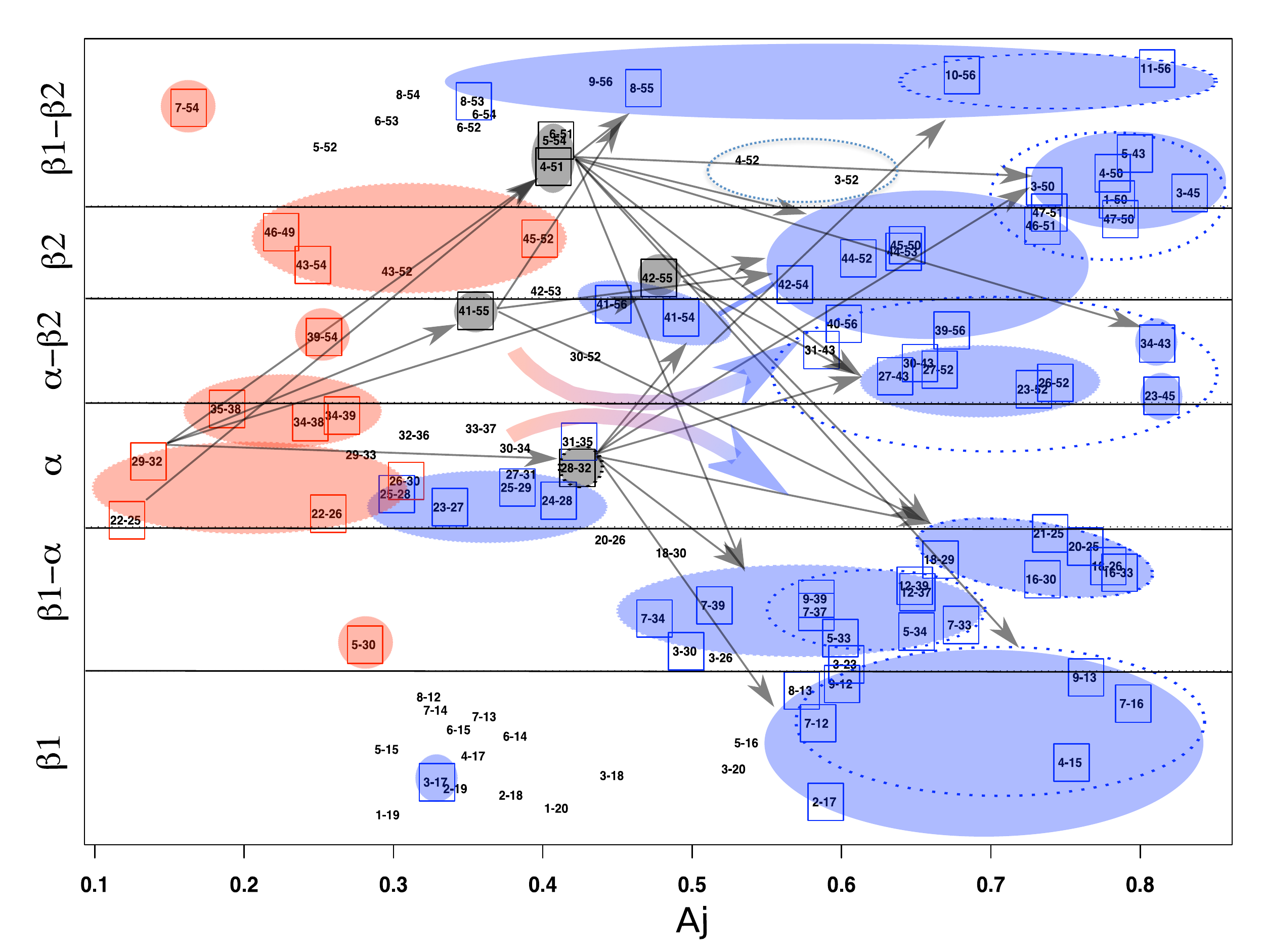}}
\caption{From the matrix $M_{ij}$ obtained from the folding simulations one can obtain the hierarchy of contact formation. The plot shows the native contacts in a plot where the horizontal axis indicate the probability $A_j$ that the jth contact is formed last, while the vertical axis indicates which regions of protein G it involves. Contacts marked with a red square are those which are always cause and never consequence of the formation of other contacts; on the contrary, contacts marked with a blue square are those which are always consequence and never cause. Contacts in a black square are  both cause and effect of other contacts. The arrows indicate a cause-effect relationship. Since the simulation shows a cause-effect relationship between essentially any red contacts to any blue contact (as suggested by the curved arrows), the arrows are explicitly indicated only for the gray contacts. The colored areas are meant to guide the eye in connection with contacts belonging to contiguous regions of the protein. Dotted ellipses indicate those contacts which are not yet formed in any of the two transition state conformations displayed in Figs. \protect\ref{strutture}(C) and (D).}
\label{freccie}
\end{figure*}

The succession of contact formation has been analyzed based on the assumption that the probability $M_{ij}$ that the formation of the $j$th contact follows the formation of the $i$th contact in the folding simulations reflects a cause--effect relationship between the two events. The resulting picture corresponds to a rather hierarchical set of contact formations. Few native contacts are always formed spontaneously, without the need of any other event to happen before. The formation of these contacts is, according to the results of the calculations summarized in Fig. \ref{freccie}, a necessary condition for the formation of the contacts at the transition state. Of notice that in this state are also formed essentially all contacts which are neither cause nor effect of other contacts (those not colored in Fig. \ref{freccie}), and few of those which are effect of the formation of other contacts (marked in blue in the figure). Contacts which are only cause of other contacts are mainly local, located within the helix and the second hairpin, while only three of them are non-local (i.e.,  5--30 between the first hairpin and the helix, 39--54 between the helix and the second-hairpin and 7--54 between the two hairpins). It is not unexpected that the residues building out these early--contacts are concentrated in the helix and in the second hairpin (cf. Figs. \ref{strutture}A and B), which are known by $\varphi$--value analysis to be structured in the transition state \cite{McC:00}.  Interestingly the non--local contact  between residues 7 and 54, linking the two terminals of the protein,  agrees with the interpretation of the effects of mutations in the first strand of protein G, according to which {\it hydrophobic residues in the first strand do make some interactions with the relatively ordered second beta--hairpin} \cite{McC:00}.

Contextualizing the above results within the framework of a two--state scenario of the folding of protein G, one can argue that the formation of the non--local contacts (5--30, 39--45 and 7--54) take place in the early stages of refolding (see also ref. \cite{Bro:98a,Bro:98b}). Within this context one can mention that, NMR spectra of the pH--denatured state of protein G highlight elements of native structure associated with the second hairpin, with the helix and with the turn of the first hairpin \cite{Sar:00}. This interpretation is consistent with results of simplified models showing the formation of local elementary structures (LES) \cite{Bro:01,Bro:04,Sut:06},  or foldons \cite{Pan:05}, in the denatured state under native conditions, and their docking into a folding nucleus \cite{Abk:94}.  Within this picture, the docking of the LES --which can be viewed as hidden incipient secondary structures but most likely lacking of a number of hydrogen boding as well as of side chains contacts as compared to the native situation-- seems also helped by the early formation of few non--local native contacts and few non--native contacts. The transition state is compatible with the docking of the LES, corresponding to a ensemble of conformations where the protein display its native topology. The end of the folding process involves the detailed packing of the buried side--chains \cite{Kus:02,Lin:04} and the stabilization through hydrogen bonds of secondary structure elements \cite{Lin:04}.

The non--native interactions seem to play two roles. First, they stabilize the helix while the system attends the formation of the native tertiary interactions. Moreover, they help the folding kinetics, attracting residues which are distant along the chain for then leaving the place to native nearby interactions. This fact suggests that evolution could have selected, at the price of protein stability, sequences not only optimizing native interactions, but also specific non--native ones able to enhance folding kinetics.

The B1 domain of Protein G folds in a time which is as fast as 5 ms \cite{Kus:94}, making the experimental characterization of the events which take place along the folding pathway, quite difficult. 
Aside from validating the model, the possibility of characterizing structurally the transition--state ensemble from {\it ab initio} calculations offers an unprecedented opportunity to complement experimental data.  Therefore, we compare our results with, on one hand, the characterization of the denatured and of the TS state, and on the other hand, with computational results obtained making use of simplified models. Also with those resulting from  studies of selected fragments of protein G.

The NMR analysis of protein G under mildly denaturant conditions carried out by  Sari and coworkers \cite{Sar:00} indicate that short--range contacts between $H_\alpha$ and $H_N$ in the turn of the first hairpin (contact 8--12), in the helix (22--25, 22--26, 29--32, 34--38) and the long--range contact 41--55 are formed in the denatured state. These contacts match remarkably well with the ``early contacts" marked in red in Fig. \ref{freccie}. Moreover, the J--couplings associated with residues 44, 49, 51 and 52 indicate that the second hairpin populates the beta region of the Ramachandran plot. Consistently with our results, this suggests that parts of the helix are formed, that the second hairpin displays a native--like topology, constrained by the contacts 39--54 and 41--55. Also, the result of the simulations is not inconsistent with the early formation of the turn in the first hairpin. 

The topology of the transition state is more similar to the native state than what experiments usually suggest. This was already noted in ref. \cite{Lin:04,Ven:01,Cal:08}, but on the basis of data--driven calculations. The difference between the transition and the native state seems to be not in the amount of native contacts, meant mainly as Van der Waals interactions, but in their degree of optimization and  in the formation of orientation--dependent H--bonds. For example, the first hairpin is quite native--like in the transition state, but it does not qualify as a beta--hairpin in terms of detailed dihedral angles and H--bonds pattern. Similarly, it is likely that the native structures present in the denatured state are not textbook secondary structures, and consequently escape characterization with traditional tools. The subtle structural difference between the transition state and the native state is likely to be the reason why simplified protein models with reduced degrees of freedom overstimate the $\varphi$--values \cite{Sut:06}. Also the reason why $\varphi$--value analysis provide a very refined microscope at the all--atom level of the transition state.

\section{Materials and methods}
\subsection{Model system}
The structure of the B1--domain domain of the IgG-binding domain of streptococcal protein G we used has pdb code 1pgb~\cite{Gal:94}. All the simulation are performed with GROMACS~\cite{Lin:01}. The interactions are described by the Amber 2003 all-atom force-field ported to Gromacs~\cite{Dua:03,Sor:05}.  The system is enclosed in a dodecahedric box of 261 ${\rm nm}^3$ with periodic boundary  conditions and solvated with 8325 SPC water molecules. The system charge was neutralized adding 4 Na$^+$ ions. Van der Waals interactions are cut-off at 1.4 ${\rm nm}$ and the long-range electrostatic interactions were calculated by the particle mesh Ewald algorithm~\cite{essman95} with a mesh spaced of 0.125~${\rm nm}$. The system evolve in the canonical ensemble, coupled with a Nos\'e-Hoover thermal bath~\cite{nose84, hoover85}.
The native state is first thermalized at 300 K for 1 ns. A 2 ns dynamics at 300 K at constant volume used to generate a reference native state ensemble.
 
\subsection{Initial conformations} 
 To generate a set of unfolded conformation we run a 26 ns long simulation at 600 K, we took 16 conformations  from the 10th to 26th ns and we thermalized them at 300K for 2 ns each. The RMSD calculated on the C$\alpha$ atoms between the 16 structures after the thermalization is between 0.8 to 1.4 nm which guarantee that the dynamics will be uncorrelated.
 
\subsection{Adiabatically biased trajectories}
The trajecories are generated by a biased molecular dynamics algorithm proposed by Marchi and Ballone~\cite{Mar:99} and applied to proteins by Paci and Karplus~\cite{Pac:99}. The driving coordinate used in the present study is the distance $d_{CM}$ of the contact map of a given protein conformation from the native contact map, introduced by Bonomi at al. in ref.~\cite{bonomi07}. This is defined as
\begin{equation}
\label{eq:dcm1}
d_{CM} = \|C-\tilde{C}\|=\left(\sum_{j>i+2}^{N}(C_{ij}-\tilde{C_{ij}})^2\right)^{1/2},
\end{equation}
were $C_{ij}$ is the i,j element of a NxN matrix defined as
\begin{equation}
\label{eq:dcm2}
C_{ij}(r_{ij}) =  \begin{cases} \frac{1-\left(\frac{r_{ij}}{r_0}\right)^p}{1-\left(\frac{r_{ij}}{r_0}\right)^q},&r_{ij} \leq r_{cut}\\
                                                     0, & r_{ij} > r_{cut}, \end{cases}
\end{equation}
$r_{ij}$ is the distance between atom $i$ and $j$ and $\tilde{C}$ is the defined on the native state.
The parameters used in these simulations are $p=6$, $q=10$, $r_0=0.75$ nm and $r_{cut}=1.23$ ${\rm nm}$.
N include all the $\alpha$ carbons and either the $\beta$ or the $\gamma$ carbons of the hydrophobic side--chains.

The biasing potential is implemented as 
\begin{equation}
V(\rho(t)) =  \begin{cases} \frac{\alpha}{2}\left(\rho(t)-\rho_m(t)\right)^2, &\rho(t)>\rho_m(t)\\
0, & \rho(t)\le\rho_m(t), \end{cases}
\end{equation}
where
\begin{equation}
\rho(t)=\left(d_{CM}(t)-0\right)^2
\end{equation}
and
\begin{equation}
\rho_m(t)=\min_{0\le\tau\le t}\rho(\tau).
\end{equation}

The first tests to chose a proper collective variable were done with a constant $\alpha$ of 40 KJ/mol, while for
the production runs the constant is set to 3 KJoule/mole. Each of the sixteen unfolded structure is evolved for 50 ns.

\subsection{Contact analysis}
A contact between two amino acids is defined if 1) there is a H--bonds between the two amino acids, that is a polar H and an O are closer than 2.5\AA~and their respective bonds are aligned with a maximum deviation of 30 deg, or 2) the minimum distance between any atom in their side--chain is less than 4\AA. Native contacts are defined if the above property holds for the average distances calculated on the native state ensemble. Having also calculated the standard fluctuations of the atom distances in the native state, a native contact is defined as stably formed once it is formed and, since then, its fluctuations do not exceed the double of those found in the native state. A non--native contact between two amino acids is defined if 1) the minimum distance between any atom is less   than 4\AA~and, 2) the mean minimum distance between any atom is more  than 5\AA~on the native state ensemble.

\begin{acknowledgments}
We are grateful to Ludovico Sutto for help and discussions. The financial support of the Italian Ministry of Scientific Research under the FIRB2003 program is acknowledged.
\end{acknowledgments}


\clearpage

\begin{thebibliography}{}
\bibitem{Han:02} Hansson T, Oostenbrink C and van Gunsteren WF (2002) Molecular dynamics simulations. {\it Curr. Opin. Struct. Biol.} 12:190--196
\bibitem{Fre:08} Freddolino P, Liu F, Gruebele M, Schulten K (2008) Ten-Microsecond Molecular Dynamics Simulation of a Fast-Folding WW Domain. {\it Biophys J}  94:L75--L77
\bibitem{Ens:07} Ensign D, Kasson P, Pande V (2007) Heterogeneity Even at the Speed Limit of Folding: Large-scale Molecular Dynamics Study of a Fast-folding Variant of the Villin Headpiece.  {\it J Mol Biol} 374:806--816 
\bibitem{Bro:98a} Sheinerman F B and Brooks C L (1998) Molecular picture of folding of a small alpha/beta protein. {\it Proc. Natl. Acad. Sci. USA} 95:1562--1567
\bibitem{Dag:01} Kazmirski S L, Wong K B , Freund S M V , Tan  Y J, Fersht A R and  Daggett V (2001) Protein folding from a highly disordered denatured state:  Folding pathway of chymotrypsin inhibitor 2 at atomic resolution. {\it Proc Natl Acad Sci} 98: 4349--4354. 
\bibitem{Mar:99} Marchi M, Ballone P (1999) Adiabatic bias molecular dynamics: A method to navigate the conformational space of complex molecular systems. {\it J Chem Phys} 110:3697-3702
\bibitem{Pac:99} Paci E, Karplus M. (1999) Forced Unfolding of Fibronectin Type 3 Modules: An Analysis by Biased Molecular Dynamics Simulations. {\it J. Mol. Biol.} 288:441--459
\bibitem{Dua:03} Duan Y, Wu C,  Chowdhury S, Lee MC, Xiong G , Zhang W, Yang R,  Cieplak P, Luo R, Lee T, Caldwell J, Wang J,  Kollman P (2003) A Point-Charge Force Field for Molecular Mechanics  Simulations of Proteins Based on Condensed-Phase Quantum Mechanical Calculations. {\it J Comp Chem} 24:1999--2012
\bibitem{Bro:01} Broglia RA, Tiana G (2001) Hierarchy of events in the folding of model proteins. {\it J Chem Phys} 114:7267--7273
\bibitem{Bro:04} Broglia RA, Tiana G, Provasi D (2004) Simple models of protein folding and of non-conventional drug design. {\it J Phys Cond Mat} 16:R111-R144
\bibitem{Kus:02} Kussell E, Shakhnovich E I (2002) Glassy Dynamics of Side-Chain Ordering in a Simple Model of Protein Folding. {\it Phys Rev Lett} 89:168101-1--168101-4
\bibitem{Bro:08} Amatori A, Tiana G, Ferkinghoff-Borg J and Broglia RA (2008) The denatured state is critical in determining the properties of model proteins designed on different folds. {\it Proteins} 70:1047--1055
\bibitem{Ser:97} Blanco FJ, Ortiz AR, Serrano L (1997) Role of nonnative interaction in the folding of the protein G B1 domain as inferred from the conformational analysis of the $\alpha$--helix fragment. {\it Fold Des} 2:123--133
\bibitem{Bro:98b} Sheinerman F B and Brooks C L (1998) Calculations on Folding of Segment B1 of 
Streptococcal Protein G {\it J. Mol. Biol.} 278:439--456
\bibitem{Cam:08} Camilloni C, Provasi D, Tiana G and Broglia RA (2008) Exploring the protein G helix free--energy surface by solute tempering metadynamics. {\it Proteins}  71:1647--1654
\bibitem{Gei:99} Geissler PL, Dellago C, and Chandler D. {\it Kinetic pathways of ion pair dissociation in water}. J. Phys. Chem. B. 1999; 103:3706--3710. 
\bibitem{paci02} Paci E, Vendruscolo M, Dobson CM, Karplus M. {\it Determination of a Transition State at Atomic Resolution from Protein Engineering Data}. J. Mol. Biol. 2002; 324:151--163
\bibitem{McC:00} McCallister E L, Alm E, Baker D (2000) Critical role of beta--hairpin formation in protein G folding, {\it Nature Struct  Biol} 7:669--673
\bibitem{Sar:00} Sari N, Alexander P, Bryan PN, Orban J (2000) Structure and dynamics of an acid-denatured protein G mutant, {\it Biochemistry} 39:965--977
\bibitem{Sut:06} Sutto L, Tiana G and Broglia RA (2006)  Sequence of events in folding mechanism: beyond the Go model. {\it Protein Sci} 15:1638--1652
\bibitem{Pan:05} Panchenko A R , Luthey-Schulten Z, Wolynes P G (2005) Foldons, protein structural modules and exons, {\it Proc. Natl. Acad. Sci. USA} 93:2008--2013
\bibitem{Abk:94} Abkevich V I, Gutin A M, Shakhnovich, E I (1994) Specific nucleus as the transition state for protein folding: evidence from the lattice model. {\it Biochemistry} 33:10026-10036
\bibitem{Lin:04} Lindorff-Larsen K, Vendruscolo M, Paci E, Dobson CM (2004) Transition states for protein folding have native topologies despite high structural variability. {\it Nat Struct Mol Biol} 11:443-449
\bibitem{Kus:94}  Kuszewski J, Clore GM, and Gronenborn AM (1994) Fast  folding of a prototypic polypeptide: 
The  immunoglobulin binding domain of streptococcal  protein G. {\it Protein Sci} 3:1945--1952
\bibitem{Ven:01} Vendruscolo M, Paci E, Dobson CM, Karplus M (2001) Three key residues form a critical contact network in a protein folding transition state. {\it Nature} 409:641-645
\bibitem{Cal:08} Calosci N, Chi C, Richter B, Camilloni C, Engstršm E, Eklund L, Travaglini-Allocatelli C, Gianni S, Vendruscolo M, Jemth P. {\it Comparison of successive transition states for folding reveals alternative early folding pathways of two homologous proteins}. Proc. Natl. Acad. USA 2008; 105:19240--19245
\bibitem{Gal:94} Gallagher T, Alexander P, Bryan P, Gilliland G. (1994) Two Crystal Structures of the B 1 Immunoglobulin-Binding Domain of Streptococcal Protein G and Comparison with NMR. {\it Biochem.} 33:4721--4729
\bibitem{Lin:01} Lindahl E, Hess B, van der Spoel D. {\it GROMACS 3.0: A package
for molecular simulation and trajectory analysis}. J. Mol. Mod. 2001; 7:306--317.
\bibitem{Sor:05} Sorin EJ and Pande VS. {\it Exploring the helix--coil transition via all--atom equilibrium
ensemble simulations}. Biophys. J. 2005; 88:2472--2493
\bibitem {essman95} Essman U, Perela L, Berkowitz ML, Darden T, Lee H, Pedersen LG. (1995)
A smooth particle mesh Ewald method. {\it J. Chem. Phys.}103:8577--8592.
\bibitem{nose84} Nos\'e S. (1984) A molecular dynamics method for simulations in the canonical ensemble. {\it Mol. Phys.} 52:255--268.
\bibitem{hoover85} Hoover WG. (1985) Canonical dynamics: equilibrium phase-space distributions. 
{\it Phys. Rev. A} 31:1695--1697.
\bibitem{bonomi07} Bonomi M, Gervasio FL, Tiana G, Provasi D, Broglia RA and Parrinello M. (2007) Insight into the Folding Inhibition of the HIV-1 Protease by a Small Peptide. {\it Biophys J.} 93: 2813--2821


\end{thebibliography}
\end{document}